\documentclass[9pt,twocolumn,twoside]{osajnl}
\usepackage[per=slash, loctolang={UK:english}, range-units=single, range-phrase=-, separate-uncertainty]{siunitx}
\usepackage{breqn}
\usepackage{physics}
\usepackage{amsmath}
\usepackage{xfrac}
\DeclareSIUnit{\pixel}{pixel}
\DeclareSIUnit{\fps}{fps}
\DeclareSIUnit{\frames}{frames}
\DeclareMathOperator{\cov}{Cov}
\journal{ao} 

\setboolean{shortarticle}{false}

\title{Object Plane Detection and Phase Retrieval from Single-Shot Holograms using Multi-Wavelength In-Line Holography}

\author[1]{Hanqing Zhang}
\author[1]{Tim Stangner}
\author[1]{Krister Wiklund}
\author[1,*]{Magnus Andersson}

\affil[1]{Department of Physics, Umeå University, 901 87 Umeå, Sweden}

\affil[*]{Corresponding author: magnus.andersson@umu.se}

\begin{abstract}
Phase retrieval and the twin-image problem in digital in-line holographic microscopy can be resolved by iterative reconstruction routines. However, recovering the phase properties of an object in a hologram needs an object plane to be chosen correctly for reconstruction. In this work, we present a novel multi-wavelength Gerchberg-Saxton algorithm to determine the object plane using single-shot holograms recorded with multiple wavelengths in an in-line holographic microscope. For micro-sized objects, we verify the object positioning capabilities of the method for various shapes and derive the phase information using synthetic and experimental data. Experimentally, we built a compact digital in-line holographic microscopy setup around a standard optical microscope with a regular RGB-CCD camera and acquire holograms of micro-spheres, \textit{E. coli} and red blood cells, that are illuminated using three lasers operating at \SI{491}{\nano\meter}, \SI{532}{\nano\meter} and \SI{633}{\nano\meter}, respectively. We demonstrate that our method provides accurate object plane detection and phase retrieval under noisy conditions, e.g., using low-contrast holograms without background normalization. This method allows for automatic positioning and phase retrieval suitable for holographic particle velocimetry, and object tracking in biophysical or colloidal research.  
\end{abstract}

\setboolean{displaycopyright}{true}

\begin{document}

\maketitle

\section{Introduction}
Three dimensional (3D) object detection and phase retrieval are two important features when investigating particle motion in colloidal research, and when studying motility of cells or bacteria \cite{Garcia-Sucerquia2006,Moon2011, Marquet2014}. To find the position of an object in 3D, phase-contrast microscopy and digital holographic microscopy (DHM) have proven powerful in combination with high-speed cameras \cite{Taute2015,Noom2014,Verrier2015,Wang2016}. Based on the experimental setup, the object position can be estimated by: fitting out-of-focus diffraction pattern with pre-recorded images at different positions \cite{Taute2015}; using the Lorenz-Mie theory of light scattering to track colloidal particles with nano-meter accuracy \cite{Verrier2015}; or calculating the center-of-mass coordinates using an discrete-dipole-approximation approach \cite{Wang2016}. These methods can achieve high accuracy for detection, however, they are limited to objects with a specific shape. To circumvent this problem, more general methods exist. For example, methods that use a focus detection criteria for either the intensity \cite{Choo2006,Zakrisson2015,Zhang2017} or phase information \cite{Xu2011} in a hologram, or methods that use the polarization-sensitive wavefront curvature \cite{Ohman2016}. However, it is a challenging task to design the criteria for object positioning with high accuracy that is also independent to the shape of the object.

To improve the accuracy in object detection, the phase information, which is accessible in most DHM techniques, can be combined with the intensity information. To retrieve the phase information, off-axis DHM contains an angled reference wave that can reconstruct the phase information without ambiguity but at a cost of reduced spatial resolution \cite{Claus2011}. In contrast, in-line DHM has higher lateral resolution compared to the off-axis setup. However, in-line DHM is more susceptible to twin image noise due to the loss of optical phase information in the detector. To resolve the twin image problem, iterative reconstruction methods have proved to have better accuracy in reconstructing phase information than those using a non-iterative approach \cite{Latychevskaia2009}. Based on iterative phase retrieval, methods have been developed that sample holograms at different: heights \cite{Greenbaum2012,Zhai2015,Rivenson2016}; angles \cite{Bianchi2017,Luo2016}; or wavelengths \cite{Xu2011,Bao2012,Noom2014}. All these methods solve the twin image problem for accurate phase retrieval. However, holograms acquired at different height often require mechanical scanning with a sample stage, increasing the complexity of experimental procedure \cite{Greenbaum2012,Rivenson2016}. In addition, acquiring several images at different height often requires the object to be immobilized. To allow for studies of objects in motion, dual-plane digital holography with multiplexed volume holographic gratings can produce single-shot holograms from different heights \cite{Zhai2015}. Besides, by using holograms recorded at different heights, angles and wavelengths all together, a propagation phasor approach can be applied to reduce the number of raw measurements \cite{Luo2016}. However, these approaches can retrieve the phase at a cost of increased complexity of the setup and alignment procedure. On the other hand, some multiple wavelength methods use relatively simple compact setups. In this case, the phase can be retrieved either by using different wavelengths that match the channels of a RGB camera \cite{Sanz2015,Farthing2017,Bianchi2017}, or tunable lasers can be used to acquire holograms at approximately ten to twenty different wavelengths, however at the cost of computational efficiency when processing the data \cite{Xu2011,Bao2012}. 

To address issues regarding high spatial resolution, fast data processing, twin image and accurate phase retrieval, and the ability to combine with existing microscope setup without introducing complicated alignment or experimental procedure to study objects in motion, we adopt multiple wavelengths in-line holography and used the retrieved phase for object detection. In this work, we present a novel multi-wavelength Gerchberg-Saxton algorithm to detect the object plane for arbitrarily shaped objects based on phase information generated from the iterative phase retrieval using multiple wavelengths in an in-line DHM. We build our DHM setup around a standard optical microscope and acquire single-shot holograms recorded with the RGB channels in a charge-coupled device (CCD) camera by illuminating objects with three wavelengths that correlate with each channel. We validate our results for accurate object plane detection and phase retrieval on synthetic and experimental data for optically semi-transparent objects such as polystyrene (PS) micro-spheres, \textit{Escherichia coli} (\textit{E. coli}) and red blood cells (RBC). Even under noisy experimental conditions, our proposed algorithm provide reliable object plane detection and phase retrieval without the need of background normalization, and therefore saving computational and experimental resources. 

\section{Materials and Methods}
\label{sec:MAM}

\subsection{Hologram Reconstruction}
\label{subsec:HologramReconstruction}

\begin{figure}[htbp]
\centering
\fbox{\includegraphics[width=\linewidth]{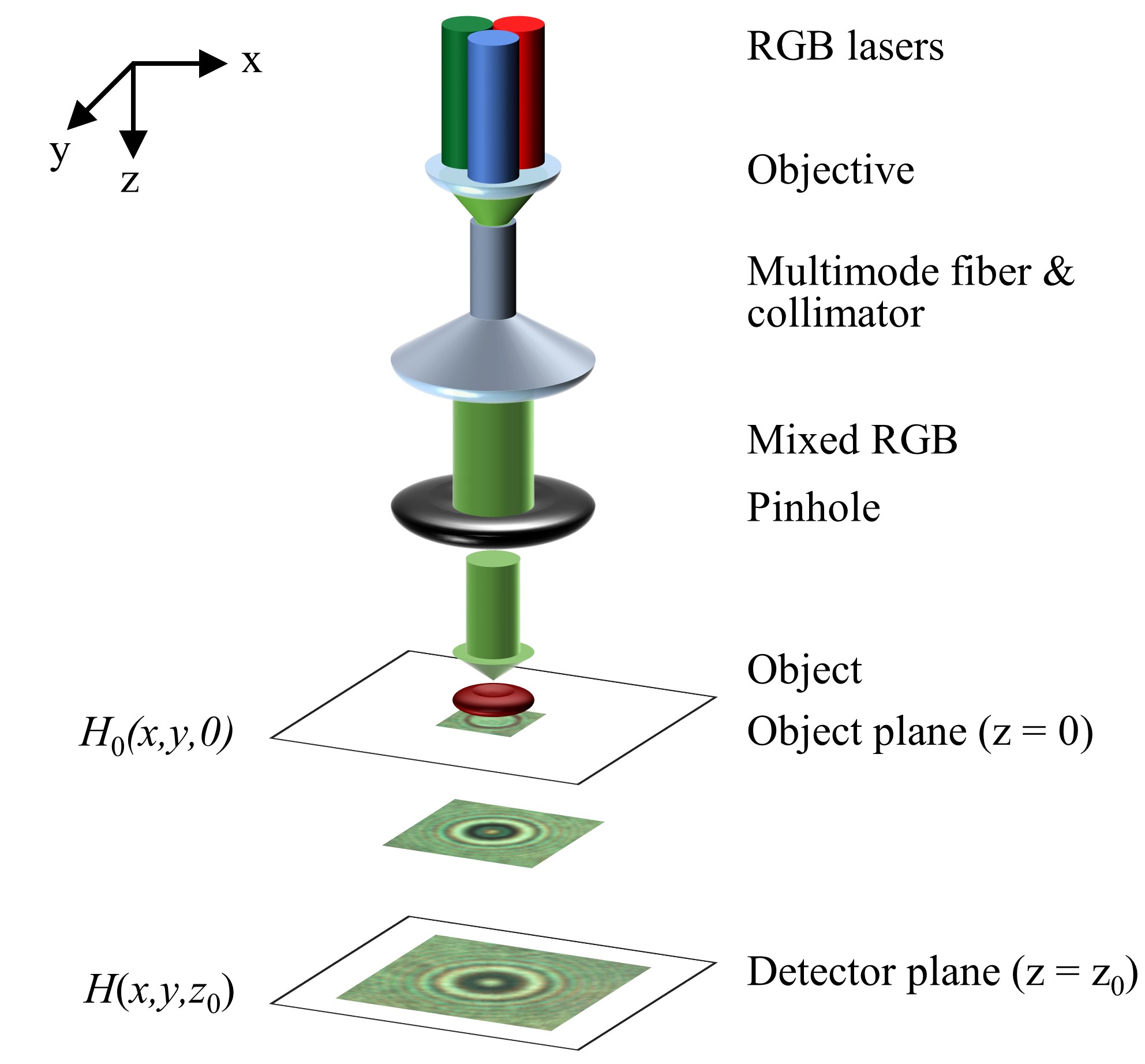}}
\caption{Experimental setup. Partial coherent light of three different wavelengths is focused via an objective into a multi-mode fiber that collimates the light. The transmitted light illuminates the object under study with plane wavefronts. To improve the contrast, we spatially filter the light after the fiber using a pinhole. We record holograms at the detector plane using a RGB-CCD camera.}
\label{fig:Figure1}
\end{figure}

In digital in-line holographic microscopy, light waves travel through an illumination system and interact with the object of interest. The scattered light from the object will interfere with non-scattered light from the source along the optical axis, which we denote the $z$-direction, resulting in a two-dimensional ($xy$) hologram in the detector plane located at $z=z_0$ (Fig. \ref{fig:Figure1}). Although the hologram $H(x,y,z_0)$ contains only intensity information, it can be related to a wavefront $U(x,y,z_0)=A(x,y)\exp(j\phi(x,y))$ containing both amplitude $A(x,y)$ and phase $\phi(x,y)$. This information is used to estimate the amplitude and phase of the object. To achieve this, we reconstruct the light propagation from the hologram at the detector back to the object using numerical methods. Since we illuminate the object with plane wavefronts we use the angular spectrum method \cite{Ratcliffe1956} in combination with the Rayleigh-Sommerfeld diffraction formula for light propagation. We apply this method to numerically reconstruct a hologram at an object plane located at $z=0$ in our coordinate system (Fig. \ref{fig:Figure1}). In this object plane, the hologram contains information of a reference wave $R(x,y,0)$ representing illumination with parallel wavefront and an object wave $O(x,y,0)$ representing the scattered light passing the object as,

\begin{dmath}
H_0(x,y,0) = |R(x,y,0)+O(x,y,0)|^2\approx R(x,y,0)(1+\tilde{O}|(x,y)),
\label{Eq:Equation1}
\end{dmath}

\noindent where $H_0$ is approximated by using the non-scattered component, $R(x,y,0)$,  and a scattered component represented by $\tilde{O}(x,y)=\exp\left(-a(x,y)\right)\cdot\exp\left(j\phi(x,y)\right)$, where $a$ is the object absorption and $\phi(x,y)$ is the phase shift caused by the object \cite{Latychevskaia2007}. In the following section, we present a multiple-wavelength Gerchberg-Saxton algorithm to retrieve the phase $\phi$ and show how this phase information can in turn be used for object plane detection.

\subsection{Object Plane Detection and Phase Retrieval: The Multi-Wavelength Gerchberg-Saxton Algorithm}
\label{subsec:IterativeMethod}

To illustrate the basic concept of object plane detection and phase retrieval using the multi-wavelength Gerchberg-Saxon algorithm, we use a two-wavelength scenario for simplicity. Note, the method can easily be expanded using several wavelengths. First, the phase $\phi(x,y)$ in the object plane (Fig. \ref{fig:Figure1}) can be estimated as,

\begin{dmath}
\phi(x,y)=\frac{2\pi}{\lambda}\left[n_{\text{o}}(\lambda)-n_{\text{m}}(\lambda)\right]h(x,y),
\label{Eq:Equation4}
\end{dmath}

\noindent where $h(x,y)$ is the object thickness, $n_{\text{o}}$ is the refractive index of the object and $n_{\text{m}}$ is the refractive index of the surrounding medium. In the following, we assume a constant, wavelength-independent refractive index for the object and its surrounding medium. With this assumption, the phases for an object at two different wavelengths $\lambda_1$ and $\lambda_2$ can then be related by,

\begin{dmath}
\frac{\phi_1(x,y)}{\phi_2(x,y)}=\frac{\lambda_2}{\lambda_1}.
\label{Eq:Equation5}
\end{dmath}

\noindent Therefore, if the phase information are correctly retrieved in the object plane, we expect a high similarity between $\phi_1(x,y)$ and the scaled $\phi_2(x,y)$. 

\begin{figure}[t]
\centering
\fbox{\includegraphics[width=\linewidth]{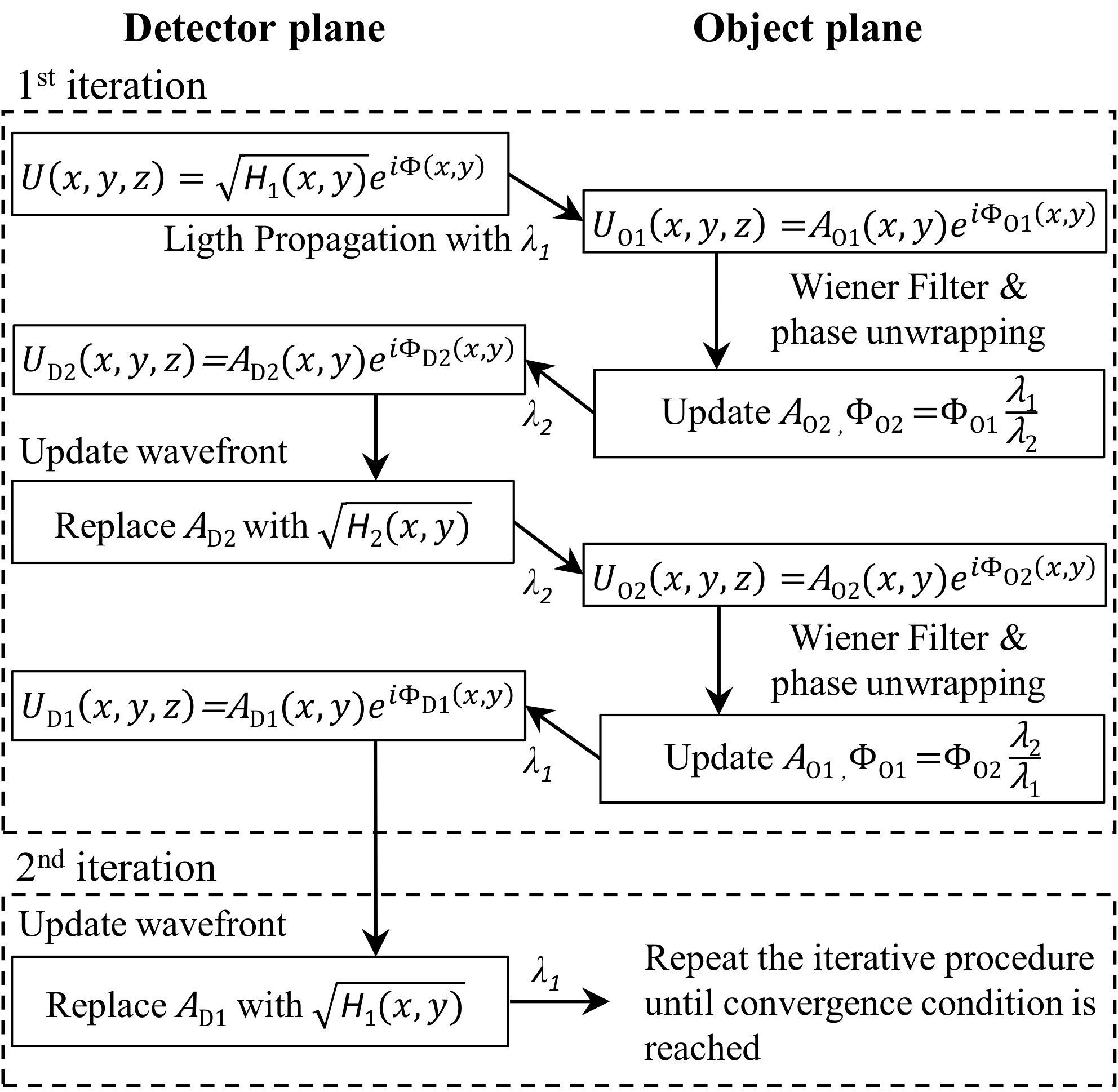}}
\caption{Work flow of the multiple-wavelength Gerchberg-Saxton algorithm simplified for two wavelengths. Arrows indicate the direction of data flow in the algorithm. For each iteration, the wavefront propagates back and forth in-between the detector plane and a potential object plane twice.} 
\label{fig:Figure2}
\end{figure}

To correctly retrieve phase information using holograms acquired at multiple wavelengths, we developed a method based on the Gerchberg-Saxton algorithm \cite{Miao1999}. At the first iteration (Fig. \ref{fig:Figure2}), we input the holograms intensity $H_1(x,y)$ and $H_2(x,y)$, acquired at different wavelengths in the detector plane. Since the detector cannot record phase information, we assign to the initial phase $\phi(x,y)$ randomized values, for example, we apply noise to all pixels from a normal distribution with zero mean, standard deviation of $0.01$, and maximum phase value of \SI{0.01}{\rad}. After setting the values for the phase, we set an initial wavefront $U(x,y,z)$ with an amplitude, defined as the squared root of $H_1(x,y)$, and the phase $\phi(x,y)$ is numerically propagated at wavelength $\lambda_1$. In a potential object plane this wavefront is denoted $U_{\text{O1}}(x,y,z)$ with amplitude $A_{\text{O1}}(x,y)$ and phase $\phi_{\text{O1}}(x,y)$. To refine the information for $U_{\text{O1}}(x,y,z)$, we apply a Wiener filter to the phase $\phi_{\text{O1}}(x,y)$ followed by an unwrapping algorithm \cite{Ghiglia1994}. Subsequently, we update $\phi_{\text{O2}}$ with the unwrapped $\phi_{\text{O1}}$ multiplied by $\lambda_1/\lambda_2$. The amplitude $A_{\text{O2}}$ is set equal to $A_{\text{O1}}$. Based on $A_{\text{O2}}$ and $\phi_{\text{O2}}$, we calculate the propagation of this wavefront at wavelength $\lambda_2$ to the detector plane. We denote the wavefront with $U_{\text{D2}}(x,y,z)$, and replace its amplitude $A_{\text{D2}}$ with the squared root of $H_2(x,y)$. A similar procedure is repeated by propagating the updated wavefront at wavelength $\lambda_2$ back to the object plane and then propagating back to the detector plane at wavelength $\lambda_1$. The second iteration is the same as the first, except $\phi(x,y)$ is updated iteratively in the process.

This algorithm can be extended to three (or more) wavelengths to utilize the RGB channels of a CCD camera. In the three wavelength case, we set the multiple wavelengths iterative phase retrieval procedure by starting from wavelength $\lambda_1$ to $\lambda_2$, $\lambda_2$ to $\lambda_3$, and going back from $\lambda_3$ to $\lambda_2$, finally reaching $\lambda_1$ to start another iteration.

In the algorithm, we iteratively update both the amplitude and phase at the detector plane and at a potential object plane. We denote the distance between the detector and a potential object plane as $L_{\text{DO}}$. To achieve stable amplitude and phase values at the potential object plane, we apply a fixed number of iterations, typically 20, until our convergence condition calculated from the sum of squared error between $H_1(x,y)$ and the square of $A_{\text{D1}}$ is below a threshold value. To find the correct object plane among potential planes, we use the input holograms and compare it with the updated amplitude reconstructed in the detector plane. If the two amplitudes are similar it is plausible that we have found the correct object plane. In the algorithm, this is realized by comparing the square of $A_{\text{D1}}$ and $A_{\text{D2}}$ with $H_1(x,y)$ and $H_2(x,y)$, respectively, using the score $S_{\text{D}}$ at each detector-object distance $L_{\text{DO}}$ as,

\begin{dmath}
S_{\text{D}}(L_{\text{DO}})=\frac{1}{n}\sum_{i=1}^n\cov\left[H_{\text{D}i},A_{\text{D}i}^2\right],
\label{Eq:Equation6}
\end{dmath}

\noindent where $\cov(H,A)$ is the covariance of matrix $H$ and $A$, and $n$ the number of wavelengths used for reconstruction. High score value corresponds to similar amplitudes, but it also indicates that the reconstruction of amplitude and phase  has a high accuracy. In addition, based on our similarity assumption in Eq. \ref{Eq:Equation5} we also check if the obtained phases from different wavelengths at a potential object plane are similar to each other by defining a similarity score $S_{\text{O}}$,

\begin{dmath}
S_O(L_{\text{DO}})=\cov\left[\phi_{\text{O1}},\phi_{\text{O2}}\right],
\label{Eq:Equation7}
\end{dmath}

\noindent where the unwrapped phase distribution $\phi_{\text{O1}}$ and $\phi_{\text{O2}}$ are related to respective wavelength, $\lambda_1$ and $\lambda_2$. For multiple wavelengths, $S_{\text{O}}$ is a sum of covariances of all 2-combinations from n wavelengths. $S_{\text{O}}$ is defined to be large when two phase distributions are similar. In our implementation, we assign a weight $w$ on $S_{\text{D}}$ and $1-w$ on $S_{\text{O}}$ , the similarity index $S$ is given as,

\begin{dmath}
S=w\norm{S_{\text{D}}}+(1-w)\norm{S_{\text{O}}}.
\label{Eq:Equation8}
\end{dmath}

\noindent Depending on the noise level and the contrast in a hologram, $w$ is tuned from 0 to 1 and $\norm{S_{\text{i}}}$ sets the values in $S_{\text{i}}$ ranging from 0 to 1. In practice, for holograms acquired at low noise level, e.g., the synthetic data containing noise only from the hologram reconstruction process, $w$ is set to 1. For noisy conditions and low contrast, e.g., the hologram of an \textit{E. coli}, $w$ is set to 0.5.

\subsection{Synthetic Data Generation}
\label{subsec:DataGenerationSynthetic}

We generate synthetic holograms for PS particles, \textit{E. coli} and RBC without background noise using a customized MATLAB routine. In the first step, we choose the simulation parameters to mimic our experimental measurements. Therefore, we set the index of refraction of the surrounding medium to $n_{\text{m}} = 1.33$ and the object's absorption to $a=0.05$. Second, we created spherical PS particles with diameter of \SI{1.04}{\micro\meter} and refractive index of $1.604$ (\SI{491}{\nano\meter}, blue), $1.598$ (\SI{532}{\nano\meter}, green) and $1.587$ (\SI{633}{\nano\meter}, red), respectively. We simplify the shape of \textit{E. coli} to be an ellipsoid with a length of \SI{3}{\micro\meter} and a width of \SI{1}{\micro\meter}. We set the refractive index of \textit{E. coli} to $n_{E. coli}=1.38$ \cite{Balaev2002}. To mimic RBCs we use a Cassini shaped model with parameter values of $a =2.2$, $b =2.25$, and $c =0.66$ representing a \SI{6.3}{\micro\meter} wide RBC-like object with an homogeneous index of refraction $n_{\text{RBC}}=1.40$ \cite{Wriedt2006,Zakrisson2015}. Third, we simulate the light propagating through the object based on Rayleigh-Sommerfeld diffraction formula and create holograms along the axial direction (Fig. \ref{fig:Figure1}, $z$-direction) by changing the distance between object and detector plane in a range from \SIrange{1}{50}{\micro\meter} with a step size $\varDelta z=\SI{1}{\micro\meter}$. At each $z$-position, we acquire one hologram. The hologram generated by the virtual detector has a size of $2001\times\SI{2001}{\pixel}$ with a \SI{88}{\nano\meter\per\pixel} (RBC) or \SI{132}{\nano\meter\per\pixel} (PS particle, \textit{E. coli}) conversion factor for hologram reconstruction. We choose the hologram size to be large enough to avoid artifacts generated from light propagation due to the shape of the hologram template.

\subsection{Sample Preparation}
\label{subsec:SamplePrep}

For DHM measurements, we use three different samples: (i) PS mono-size particles: \SI{1.040 \pm 0.022}{\micro\meter} (average diameter $\pm$ standard deviation (SD), Lot No. 15879, Duke Scientific Corp., \SI{4}{\%w/v}, (ii) \textit{E. coli} (C600), (iii) RBCs from a healthy voluntary donor. Each sample is diluted $1:20000$ in phosphate buffered saline (PBS, 1x, pH 7.4). 

To carry out DHM experiments, we use a custom-made sandwich chamber, consisting of two coverslips (lower coverslip: no. 1, Knittel Glass, \SI{60x24}{\milli\meter}; upper coverslip: no. 1, Knittel Glass, \SI{20x20}{\milli\meter}) separated by a layer of vacuum grease (Dow Corning). In detail, we first functionalize the lower coverslip with \SI{0.01}{\%} poly-L-lysine (catalog no. P4832, Sigma Aldrich) by adding a \SI{50}{\micro\liter} droplet to the coverslip center and heat the sample for \SI{60}{\min} at \SI{60}{\celsius} in an oven. Next, we add a vacuum grease ring around the poly-L-lysine coated area and pipette \SI{10}{\micro\liter} of prepared sample solution onto this region. Next, we seal the chamber by placing the upper coverslip on top of the vacuum grease. After incubating the measurement chamber for \SI{30}{\minute} (RBC) or over night (PS particles, \textit{E. coli}) at room temperature, the sample objects settle down and immobilize to the bottom coverslip.

\subsection{RGB Digital In-Line Holographic Microscopy}
\label{subsec:Setup}

We build the multi-wavelength DHM setup (Fig. \ref{fig:Figure1}) around an Olympus IX70 inverted microscope, normally used for optical tweezers experiments \cite{Fallman2004}. For sample illumination, we use a red (\SI{633}{\nano\meter}, HeNe-laser, 1137 Uniphase, Manteca), a green (\SI{532}{\nano\meter}, Samba\texttrademark, 05-01 Series, Cobolt AB) and a blue laser (\SI{491}{\nano\meter}, Calypso\texttrademark, 04-01 Series, Cobolt AB). To reduce speckle noise during image acquisition, we focus all three lasers on a rotating ground glass \cite{Stangner2017}. We collect the scattered light from the rotating ground glass using a plan achromat objective ($10\times$, RMS10x, Thorlabs) and focus the light into a multi-mode fiber (M76L02, Thorlabs) using a fiber launch system (MBT613D, Thorlabs). The multi-mode fiber scrambles the incoming light. To achieve better contrast in the detector plane, we spatially filter the collimated fiber output using a pinhole (P300S, ST1XY-D, Thorlabs). To ensure illumination with plane wavefronts, we position the pinhole $\SI{30}{\milli\meter}$ (corresponding to $\approx 4500\times\lambda_{\text{red}}$) above to sample.

We mount the prepared sample chamber onto a \textit{xyz}-piezo stage which can be positioned in three dimensions over a range of \SI{100}{\micro\meter} with nanometer accuracy using piezo actuators (P-561.3CD, Physik Instrumente). Subsequently, we image the object under study using an oil-immersion objective (PlanApo $60\times/1.40$ Oil, $\infty/0.17$, Olympus) and record its hologram in the detector plane using a RGB camera (MotionBLITZ EoSens Cube 7, Mikrotron, pixel size of $8\times\SI{8}{\micro\meter}$) operating at a shutter time of \SI{100}{\milli\second}. We acquire images by MotionBLITZDirector2 software with an image size of $1696\times\SI{1710}{\pixel}$ and a frame rate of \SI{10}{\fps}. By using the $60\times$ objective we obtain an optical resolution of the microscopy system of \SI{132\pm2}{\nano\meter\per\pixel}. To achieve optimal phase information in the acquired hologram for RBCs, we use an additional $1.5\times$ built-in magnification from the microscope, resulting in a conversion factor of \SI{88\pm2}{\nano\meter\per\pixel}. The whole setup is built in a temperature controlled room at \SI{23\pm1}{\celsius} to ensure long-term stability and to reduce thermal drift effects.

\section{Results and Discussion}
\label{sec:ResultsDiscussion}

\subsection{Validation of Object Plane Detection}
\label{subsec:ValidationObjectPlane}

\begin{figure}[hbtp]
\centering
\fbox{\includegraphics[width=\linewidth]{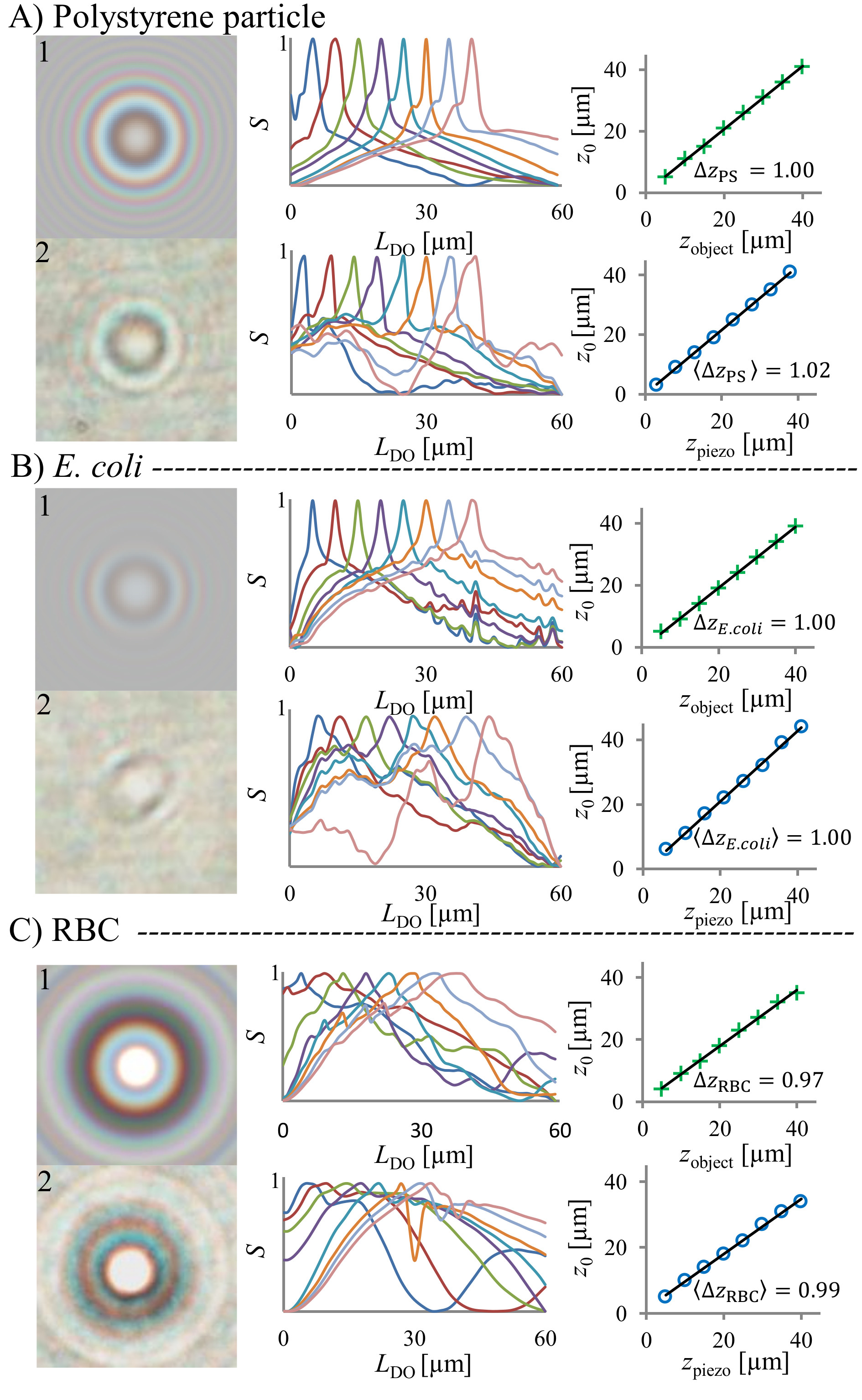}}
\caption{Object plane detection using the multi-wavelength Gerchberg-Saxton algorithm. \textit{Left column.} Synthetic and measured RGB holograms for a \SI{1}{\micro\meter} PS particle, an \textit{E. coli}, and a RBC acquired \SI{10}{\micro\meter} away from the detector plane. \textit{Middle column.} Object plane detection using the similarity index $S$ in dependence of the searching range $L_{\text{DO}}$ with \SI{1}{\micro\meter} searching steps along the optical axis. Each peak corresponds to an object plane position $z_0$. \textit{Right column.} Comparison between detected object plane position $z_0$ and its respective ground truth ($z_{\text{object}}$ or $z_{\text{piezo}}$). In each case, we obtain a linear relation with a slope value of $\varDelta z\approx 1$ (coefficient of determination $R^2=0.99$), proving that our algorithm can accurately find the object plane for objects with various shape, size and index of refraction.}
\label{fig:Figure3}
\end{figure}

To validate the capability of our multiple-wavelength Gerchberg-Saxon algorithm to find the object plane position for objects of different shapes and index of refraction, we first analyze synthetic holograms for PS particles, \textit{E. coli} cells, and RBCs (Fig. \ref{fig:Figure3} A1-C1). To find the correct object plane position relative to the detector $z_0$, our algorithm calculates the similarity index $S$ (Eq. \ref{Eq:Equation8}) along the optical axis using the reconstructed, unwrapped phase information (Fig. \ref{fig:Figure3}, middle column). For that purpose, we choose manually the searching range along the optical axis $L_{\text{DO}}$ to be \SIrange{0}{60}{\micro\meter} with a step size of \SI{1}{\micro\meter} and use 20 iterations for reconstruction in the algorithm. After evaluating holograms at various distances from the detector, we determine the relative object plane position $z_0$ by finding the maximum value of the similarity index $S$ along the optical axis and plot this value against its ground truth from the simulation (Fig. \ref{fig:Figure3}, right column, green crosses). For better visibility, we plot $z_0$ only for discrete steps of \SI{5}{\micro\meter}.

For all three test objects, we find a linear relationship between $z_0$ and its ground truth value $z_{\text{object}}$ from the simulation (Fig. \ref{fig:Figure3}, right column, black lines). From the slope of the linear regression $\varDelta z$, we obtain a height difference between two subsequent object planes of $\varDelta z_{\text{PS, sim}}=\SI{1.00\pm 0.00}{\micro\meter}$ (PS) and $\varDelta z_{E. coli\text{, sim}}=\SI{1.00\pm 0.00}{\micro\meter}$ (\textit{E. coli}), showing perfect agreement with the step size parameter set in our simulations. For the RBC our algorithm estimates the step size parameter with a slope of the linear regression $\varDelta z=\SI{1.00\pm 0.00}{\micro\meter}$ in a detector-object distance range from \SIrange{5}{10}{\micro\meter}, and $z=\SI{0.97\pm 0.02}{\micro\meter}$ in a detector-object range from \SIrange{5}{40}{\micro\meter}. We attribute this underestimation to the number of iterations in combination with the phase unwrapping during the iterative phase retrieval process. In case the iterative reconstruction process generates a phase delay bigger than $2\pi$ for an optical thick object, the phase can be resolved by the unwrapping algorithms. However, if the iteration number is not sufficiently large to resolve the phase information, phase unwrapping results become error-prone. As a consequence, the signal-to-noise ratio for the similarity index $S$ is low (Fig. \ref{fig:Figure3}, RBC, middle column), leading to a less accurate object plane detection. To further improve the detection, we recommend to use a higher optical magnification for simulations and experiments, since higher spatial resolution in the acquired hologram ensures accurate phase retrieval, minimizing errors from phase unwrapping. For this reason, we use a $90\times$ magnification for simulations and experiments involving RBCs (section \ref{sec:MAM}\ref{subsec:DataGenerationSynthetic} \& \ref{sec:MAM}\ref{subsec:Setup}).

To validate the performance using experimental data, we acquire holograms for PS particles, \textit{E. coli} cells and RBCs as specified in section \ref{sec:MAM}\ref{subsec:Setup}. In detail, we first focus on the immobilized sample objects on the coverslip ($z=0$). Next, we perform a $1D$ scan along the optical axis, covering a range from \SIrange{0}{50}{\micro\meter} with a step size of $\varDelta z_{\text{piezo}}=\SI{1}{\micro\meter}$. At each height, we acquire at least one hologram (Fig. \ref{fig:Figure3}, A2-C2). From these holograms, we reconstruct the intensity and phase information, estimate the similarity index $S$ along the optical axis and determine $z_0$. In total, we analyze 15 PS particles, 12 \textit{E. coli} cells and 12 RBCs to achieve statistically reliable results. 

For PS particles, \textit{E. coli} cells and RBC, we find over the entire scan range a linear relationship between $z_0$ and piezo stage position $z_{\text{piezo}}$ (Fig. \ref{fig:Figure3}, right column, open blue spheres). From the slope of the respective linear regressions, we determine the mean height difference between two subsequent object planes to: $\varDelta z_{\text{PS, exp}}=\SI{1.02\pm 0.01}{\micro\meter}$ (PS), $\varDelta z_{E. coli\text{, exp}}=\SI{1.00\pm 0.04}{\micro\meter}$ (\textit{E. coli}) and, $\varDelta z_{\text{RBC}}=\SI{0.99\pm 0.09}{\micro\meter}$ (RBC), matching the step size of the piezo stage $\varDelta z_{\text{piezo}}$.

\subsection{Evaluation of Phase Retrieval}
\label{subsec:ValidationPhase}

After proving that our multi-wavelength Gerchberg-Saxton algorithm correctly detects the object plane for objects of different shape and index of refraction, we demonstrate its ability to extract accurate phase information from a noisy, single-shot hologram. 

We first analyze the synthetic holograms for PS particles, \textit{E. coli} and RBCs using our proposed method. We retrieve phase information using single holograms at a fixed object-detector distance $z_0=\SI{10}{\micro\meter}$ for three wavelengths and compare the obtained values with the ground truth generated from the simulation (section \ref{sec:MAM}\ref{subsec:DataGenerationSynthetic}). For PS particles and \textit{E. coli} cells, our multi-wavelength Gerchberg-Saxton algorithm achieves a \SI{99}{\%} agreement with the ground truth after 100 iteration. However, for RBCs we achieve the same accuracy for after 500 iterations due to their size and complex structure.

Second, we acquire a single hologram at $z_0=\SI{10}{\micro\meter}$ using the RGB-DHM setup (section \ref{sec:MAM}\ref{subsec:Setup}), and we extract the phase information for PS particles, \textit{E. coli} cells and RBCs and compare these values to the reference phase from the simulation data. To further verify our multi-wavelength Gerchberg-Saxton algorithm, we compare the obtained phase values to the results from a well-established multi-height method \cite{Greenbaum2012,Rivenson2018}. This method estimates the phase information by analyzing single wavelength holograms acquired at various heights along the optical axis. For that purpose, we measure holograms with their detector-object distance ranging from the object focus up to \SI{40}{\micro\meter} with a step size of $\varDelta z=\SI{5}{\micro\meter}$ at a fixed wavelength of $\lambda_{\text{blue}}=\SI{491}{\nano\meter}$. 

\begin{figure}[hbpt]
\centering
\fbox{\includegraphics[width=\linewidth]{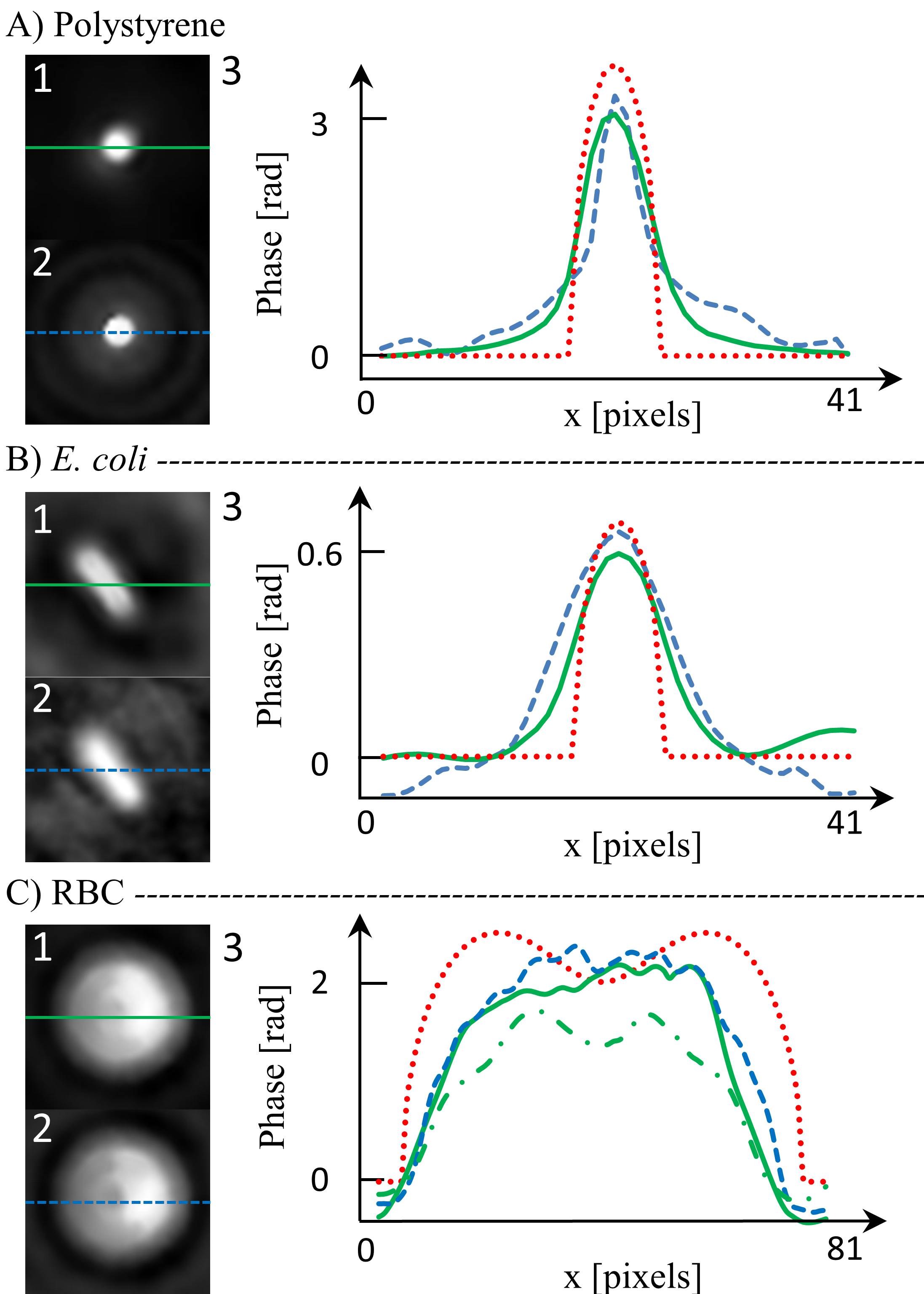}}
\caption{Comparison between phase retrieval results for A) PS particle, B) \textit{E. coli} and C) RBC generated by using 1. multi-wavelength Gerchberg-Saxton method and 2. mutli-height method. 3. The sampling of phase profile along the green solid lines and dashed blue lines corresponds to our method and the multi-height method, respectively. Note that both methods used 100 iterations to obtain results for PS particle and \textit{E. coli} data. As for RBC data, we used 200 iterations with our method. The green dashed dot line in C3 represents results from our method using 100 iterations. We demonstrate here only a single image channel corresponding to wavelength at \SI{491}{\nano\meter}. The red dotted line represents the reference phase distribution from simulations for a PS particle, an \textit{E. coli} cell and one RBC with maximum phase value of $\SI{3.51}{\radian}$, $\SI{0.64}{\radian}$ and $\SI{2.36}{\radian}$, respectively. }
\label{fig:Figure4}
\end{figure}

For PS particles, our proposed algorithm reveals a maximum phase value $\phi_{\text{PS},\lambda}=\SI{2.89\pm 0.41}{\radian}$, while the multi-height method produces $\phi_{\text{PS,h}}=\SI{2.97\pm 0.39}{\radian}$ (Fig. \ref{fig:Figure4}A). In total, we analyze 15 samples to get statistically reliable results and we use for each sample 100 iterations to extract phase values. Compared to the phase value obtained in simulations, the two methods underestimate the maximum phase by $\approx \SI{18}{\%}$ (multi-wavelength) and $\approx \SI{15}{\%}$ (multi-height), respectively. We attribute this deviation to noise in the acquired holograms and to deviations in diameter of the used PS particles. However, more importantly, the result from our multi-wavelength Gerchberg-Saxton algorithm differs only \SI{3}{\%} from the multi-height approach, despite analyzing only a single hologram acquired at three wavelengths.

Next, we evaluate the phase information of 12 \textit{E. coli} cells and 12 RBCs (Fig. \ref{fig:Figure4}B,C) and compare the obtained values to the reference phase from simulations. However, due to variations in the object's shape, orientation and tilt, the simulation value might differ and acts only as reference. Nonetheless, the reconstructed phase information using our multi-wavelength approach (Fig. \ref{fig:Figure4}, solid green line) and the multi-height method (Fig. \ref{fig:Figure4}, dashed blue line) reproduce the phase distribution for \textit{E. coli} similar to the reference phase, by deviating in average $\approx \SI{10}{\%}$ (multi-wavelength) and $\approx \SI{6}{\%}$ (multi-height) from the simulated maximum phase value. For 100 iterations the phase difference between both methods are only \SI{4}{\%}. For RBC data, we initially use 100 algorithm iterations and obtain a maximum phase $\phi_{\text{RBC}}=\SI{2.97\pm 0.39}{\radian}$ (multi-wavelength) and $\phi_{\text{RBC}}=\SI{2.49\pm 0.65}{\radian}$ (multi-height). However, in this case our proposed algorithm underestimates the reference phase from simulation by $\approx \SI{20}{\%}$ and the phase from the multi-height approach by $\approx \SI{24}{\%}$ (Fig. \ref{fig:Figure4}C3, green dash dotted line). This discrepancy can be resolved by doubling the iteration number during phase retrieval, resulting in a maximum phase of $\phi_{\text{RBC}}=\SI{2.37\pm 0.72}{\radian}$ (Fig. \ref{fig:Figure4}C3, green solid line).

With these results, we want to emphasis that we achieve high object plane detection accuracy of \textit{E.coli} and RBCs (Fig. \ref{fig:Figure3} B2 and C2, right column) by using only 20 iterations, showing the robustness of our proposed multi-wavelength Gerchberg-Saxton algorithm. However, if higher accuracy in object plane detection is desired, the number of iterations for phase retrieval can be increased.

\section{Conclusion}
In this work, we developed a novel multi-wavelength Gerchberg-Saxton algorithm to detect the position of an object using single-shot holograms acquired in an in-line DHM setup. We validated our method using both synthetic and experimental data of micro-sized PS micro-spheres, \textit{E. coli} and RBCs. We demonstrated that the object plane detection produce the same absolute positions for the tested objects of different shapes. We showed our method was capable of conducting object plane detection and phase retrieval using noisy raw holograms without normalization.

Depending on the application, our method can be applied to track object positions using in-line DHM with high speed recording, or be used to automatically retrieve phase information using a DHM setup built around a conventional optical microscope system. 

\section{Funding Information}

T.S. acknowledges financial support from the German Research Foundation (DFG) via a postdoctoral fellowship. This work was supported by Kempestiftelserna to M.A.





\end{document}